# Linking energy loss in soft adhesion to surface roughness

Siddhesh Dalvi[1*], Abhijeet Gujrati[2*], Subarna R. Khanal[2], Lars Pastewka[3], Ali Dhinojwala[1#], Tevis D. B. Jacobs[2#]

[1]Department of Polymer Science, The University of Akron, Akron, Ohio, 44325, United States
[2]Department of Mechanical Engineering and Materials Science, University of Pittsburgh, Pittsburgh, Pennsylvania, 15261, United States
[3]Department of Microsystems Engineering, University of Freiburg, 79110 Freiburg, Germany

*these two authors contributed equally
#corresponding authors

## Abstract

A mechanistic understanding of adhesion in soft materials is critical in the fields of transportation (tires, gaskets, seals), biomaterials, micro-contact printing, and soft robotics. Measurements have long demonstrated that the apparent work of adhesion coming into contact is consistently lower than the intrinsic work of adhesion for the materials, and that there is adhesion hysteresis during separation, commonly explained by viscoelastic dissipation. Still lacking is a quantitative experimentally validated link between adhesion and measured topography. Here, we used *in situ* measurements of contact size to investigate the adhesion behavior of soft elastic polydimethylsiloxane (PDMS) hemispheres (modulus ranging from 0.7 to 10 MPa) on four different polycrystalline diamond substrates with topography characterized across eight orders of magnitude, including down to the Ångström-scale. The results show that the reduction in apparent work of adhesion is equal to the energy required to achieve conformal contact. Further, the energy loss during contact and removal is equal to the product of the intrinsic work of adhesion and the true contact area. These findings provide a simple mechanism to quantitatively link the widely observed adhesion hysteresis to roughness rather than viscoelastic dissipation.

## Statement of Significance

Despite its practical importance, the adhesion of soft materials to rough surfaces is not understood. Rough surfaces show adhesion hysteresis: a difference in adhesion behavior between loading and separation. This behavior is typically attributed (often without evidence) to viscoelastic energy dissipation and cannot at present be quantitatively linked to surface topography. This work uses *in situ* measurements of contact between soft elastomers of varying stiffness against hard substrates, whose topography was characterized down to the atomic scale. Results reveal the direct relationship between Griffith fracture and the detachment of adhesive contacts and suggest a quantitative model that describes both the loading and separation behavior. The findings will guide the development of reversible adhesion for soft robotics, biomaterials, and pick-and-place manufacturing.

## Introduction

Many natural and engineering processes—such as a human picking up an object, a gecko climbing trees, or a tire gripping the road—require enough adhesion and friction to achieve the task, while maintaining the ability to release the surface afterward (1–3). All natural and manmade surfaces contain roughness at some scales, and this roughness strongly affects adhesion (4–6). Therefore, fundamental understanding of the reversible adhesion of a soft material to a rough surface is a grand scientific challenge, with significant technological applications. For instance, pick-and-place techniques in manufacturing are used from large-scale factories (7) all the way down to nanoscale transfer printing (8). Biomedical devices must securely adhere to skin when measuring vital signs or delivering drugs, but then must be removed without pain for disposal or reuse (9). Tires, seals, and gaskets are used extensively in vehicles and industrial machinery (3). And finally, there has been significant recent progress in the field of soft robotics, with the goal of creating machines that will be able to manipulate objects like the human hand or climb walls like geckos (10).

Our understanding of adhesive contact between smooth soft elastic materials was elegantly resolved in a seminal paper in 1971, where Johnson, Kendall, and Roberts (JKR) showed that the contact area under applied load is larger than predicted by the classic Hertz model (11, 12). However, the presence of surface roughness significantly alters the contact behavior. As a rough contact is loaded, it obeys the *trends* of the JKR model, but the measured apparent work of adhesion $W_{app}$ is significantly lower than the intrinsic value $W_{int}$; the latter is a thermodynamic parameter that depends on intermolecular interactions between the materials (13). Upon retraction, adhesion hysteresis is observed on rough surfaces, where the behavior deviates significantly from that of loading and from the JKR predictions. If the JKR formalism is applied, one calculates a work of adhesion (for retraction) that is much larger than $W_{int}$ and may not have thermodynamic significance (14–18).

Even though all practical surfaces are rough, quantifying the loss of adhesion due to roughness has remained a challenge. One widely-used category of models describes rough surfaces as an array of individual contacting bumps (asperities) of a certain size (19–21), where the surface properties are computed from the collective behavior of the individual asperities. However, these models focus only on a single size scale of roughness, whereas most natural and engineering surfaces are rough over many length scales (22, 23). To address the multi-scale nature of roughness, Persson developed a set of continuum mechanics models to describe soft-material adhesion at rough contacts as a function of the power spectral density (PSD) (24–26). The PSD, $C$, is a mathematical tool for separating contributions to topography from different length scales λ, and is commonly represented as a function of wavevector q = 2π/λ. In particular, under the assumption that the soft material fully conforms to the roughness of the hard material, and by assuming that the materials behave linear elastically, one such model (24) predicts how the intrinsic work of adhesion $W_{int}$ can be replaced by an apparent value $W_{app}$ that depends on material parameters and surface roughness.

The key remaining challenge for the experimental validation and practical application of these recent contact theories has been the experimental measurement of surface topography across all size scales. Surface roughness exists down to the atomic scale, and these smallest scales have been shown to be critically important for contact and adhesion (3, 6, 27, 28). Yet the conventional techniques for measuring surface topography—such as stylus or optical profilometry and atomic force microscopy—are incapable of measuring roughness down to the nanoscale (29, 30). Furthermore, because surface roughness exists over many length scales, no single technique is capable of characterizing a surface completely (30). The novelty of the present investigation lies in the combination of well-controlled adhesion measurements with complete characterization of topography across all scales, spanning from millimeters to Ångströms. This all-scale characterization eliminates the assumptions (30, 31) that are typically required for describing a surface beyond the bounds of measurement (such as self-similarity or self-affinity), thus enabling unprecedented scientific insight into the nature of rough-surface adhesion. Without such comprehensive topography measurements, the assumptions and accuracy of soft-material contact theories remain untested.

While the aforementioned mechanics models describe the behavior of a material under load, they do not predict the adhesion hysteresis, the difference in behavior between loading and separation. Instead, the increase in adhesion energy upon retraction is often attributed (sometimes without evidence) to velocity-dependent dissipation of energy due to bulk viscoelasticity (32–34). However, roughness-induced adhesion hysteresis is still observed even for systems that show no evidence of viscoelasticity on smooth surfaces (35, 36). Furthermore, it may not even be appropriate to apply an equilibrium-based theoretical model (such as JKR for smooth surfaces or Persson's model for rough surfaces) to the non-equilibrium separation behavior (37, 38). Thus, our current understanding of adhesion hysteresis is incomplete. Here, we investigate the origins of energy loss in order to demonstrate the fundamental contribution of surface roughness.

## Results and Discussion

To understand the dependence of adhesion on roughness, we performed *in situ* measurements of the load-dependent contact of sixteen different combinations of soft spheres and rough substrates. We have chosen PDMS as our elastomer and synthetically grown hydrogen-terminated diamond as the hard rough substrates because both have low surface energies. We wanted to avoid adhesion hysteresis due to interfacial bonding (for example, PDMS in contact with silica surfaces) (39, 40);

therefore, low-energy materials were chosen (41) to focus specifically on the adhesion hysteresis that arises due to surface topography.

We used a recently developed approach (29) to characterize the surface topography of four different nanodiamond substrates across eight orders of magnitude of size scale, including down to the Ångström-scale (Fig. 1). More than 50 individual topography measurements were made for each substrate (see Methods and SI Appendix, Section 1) using transmission electron microscopy, atomic force microscopy, and stylus profilometry. Results were combined to create a single comprehensive power spectral density describing each surface.

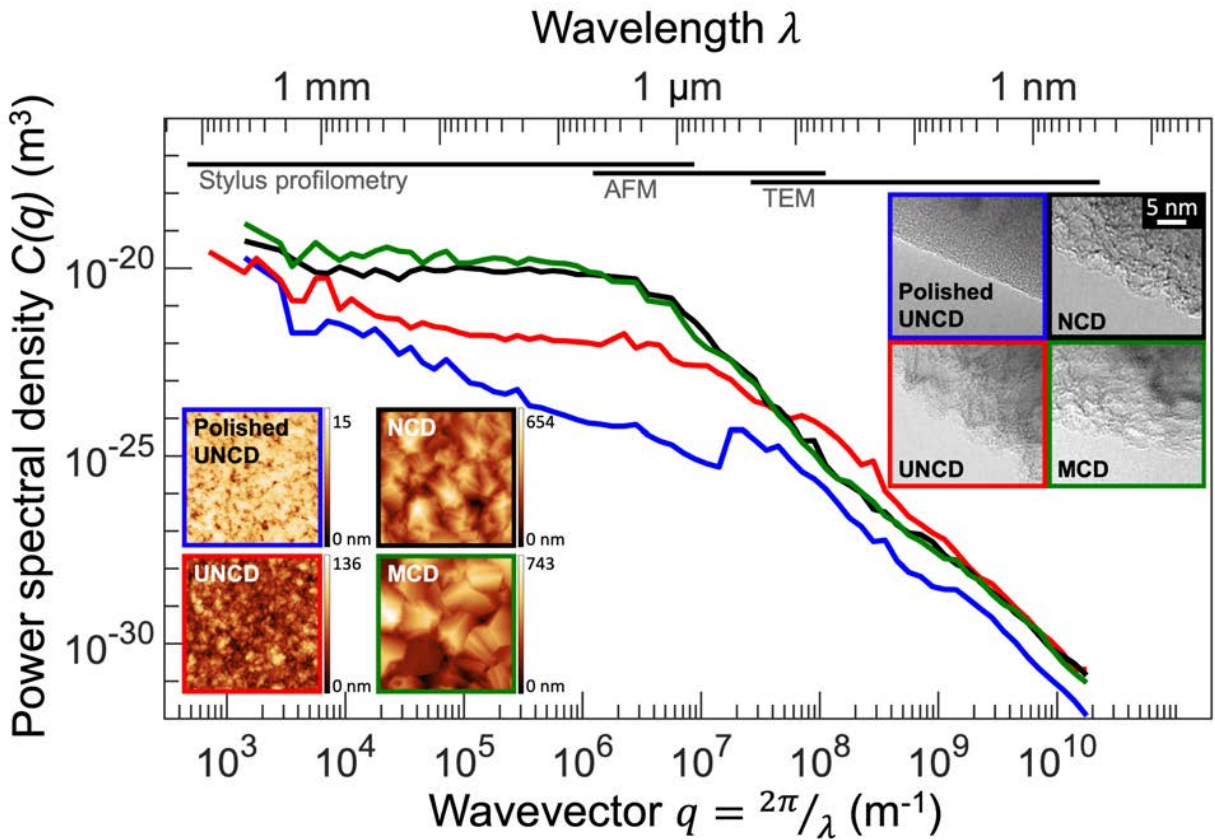

**Figure 1:** Comprehensive topography characterization for four rough nanodiamond surfaces. The surface topography was measured using a multi-resolution approach that combines transmission electron microscopy (TEM), atomic force microscopy (AFM), and stylus profilometry. Regions of applicability of each technique are indicated with horizontal bars, and are delineated more specifically in Fig. S2. The nanodiamond surfaces are designated using the following nomenclature: ultrananocrystalline diamond (UNCD) is shown in red; nanocrystalline diamond (NCD) in black; microcrystalline diamond (MCD) in green, and a polished form of UNCD (polished UNCD) in blue. AFM images (of 5-micron lateral size) are shown in the left inset; TEM images are shown in the right inset. More than 50 measurements for each surface are combined using the power spectral density, which reveals the contribution to overall roughness from different length scales (wavelengths). These comprehensive descriptions of surface topography enable the determination of true surface area and stored mechanical energy due to the topography, which are necessary to understand adhesion.

Four types of soft, elastic PDMS hemispheres were synthesized following the methods from Refs. (34, 42, 43) with elastic moduli ranging from 0.7 to 10 MPa. The PDMS hemispheres were loaded under displacement control to a maximum load of 1 mN before unloading to separation. (The synthesis and testing are described in Methods and SI Appendix, Section 2). Real-time measurements were made of contact radius, load, and displacement, as shown in Fig. 2.

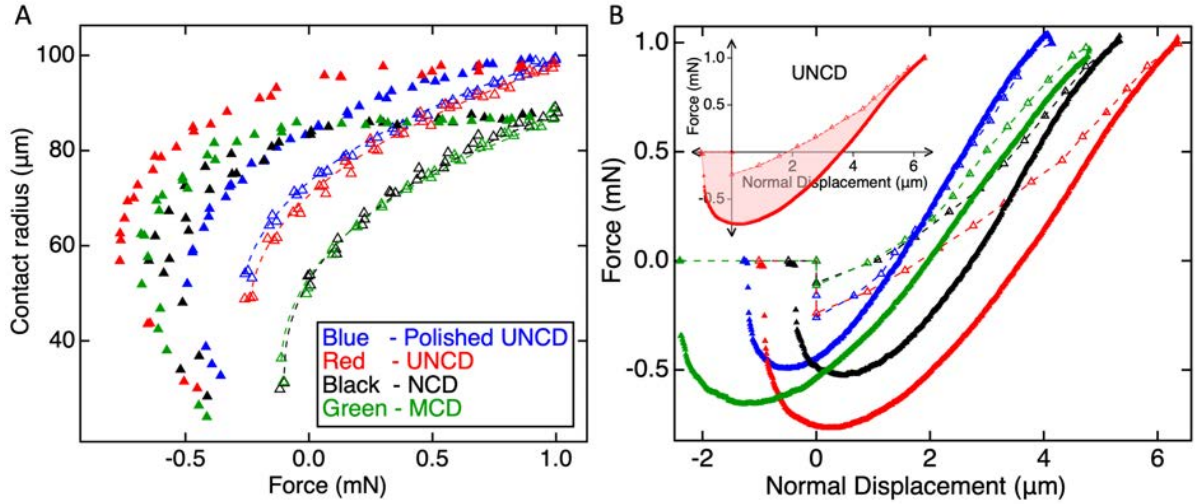

**Figure 2:** Adhesion measurements during approach and retraction. Loading and adhesion tests were performed with ultra-smooth PDMS hemispheres of varying stiffness from 0.7 to 10 MPa. Representative curves from one material (with E=1.9 MPa) are presented in this figure, with those of other materials shown in Fig. S5. The load-dependent contact radius (A) was measured using *in situ* optical microscopy. The apparent work of adhesion upon approach $W_{app}$ was extracted by fitting the loading data (hollow points) using the JKR model (dashed lines). The force-displacement curves (B) were used to calculate the energy loss $E_{loss}$ during contact by performing a closed-circuit integral (inset). Both approach and retraction experiments were conducted at a very low speed, 60 nm/s.

The apparent work of adhesion during approach $W_{app}$ is extracted by using the JKR model to fit the measured contact radius *a* as a function of load *F* (12):

$$a = \left[ \frac{3R}{4E^*} \left( F + 3\pi R W_{app} + \sqrt{6\pi R F W_{app} + \left(3\pi R W_{app}\right)^2} \right) \right]^{1/3} \quad (1)$$

where $R$ is the radius of the hemispherical lens and the effective modulus $E^*$ is defined as $1/E^* = (1 - \nu_{sphere}^2)/E_{sphere} + (1 - \nu_{substrate}^2)/E_{substrate}$, $E$ is Young's modulus and $\nu$ the Poisson ratio. This yields a different value of apparent work of adhesion for each of the sixteen contacts. The surface chemistry of the PDMS and the nanodiamond is expected to be similar in all cases, therefore all contacts should have approximately the same value of $W_{int}$. Before testing the hemispheres with rough surfaces, they were tested against a smooth silicon wafer coated with a low-surface energy octadecyltrichlorosilane (OTS) monolayer to verify that there is negligible adhesion hysteresis due to viscoelasticity (Fig. S6).

To analyze the dependence of $W_{app}$ on modulus and multi-scale surface topography, we use a model of conformal contact, based on Persson and Tosatti (24). Those authors postulated that the product of $W_{app}$ and $A_{app}$ (the apparent or projected area) is given by a balance of adhesive energy and stored elastic energy $U_{el}$:

$$W_{app}A_{app} = W_{int}A_{true} - U_{el} \tag{2}$$

with $W_{int} = \gamma_1 + \gamma_2 - \gamma_{12}$, where $\gamma_1$ and $\gamma_2$ are the surface energies of the soft and hard surfaces, respectively, and $\gamma_{12}$ is the interfacial energy between them. The term $A_{true}$ is the true surface area of the rough hard surface. However, Eq. 2 makes two important assumptions that must be addressed: it neglects the change in area of the soft elastomer surface from $A_{app}$ to $A_{true}$ upon contact; and it assumes that the surface energy of the soft material is independent of strain. These two assumptions can be corrected by modifying the energy balance to explicitly include the work done in increasing the surface area of the elastomer.

The Persson-Tosatti energy balance implicitly implies that the area of the PDMS surface does not change. While this may be valid for small-slope surfaces, in the more general case the area will increase from $A_{app}$ to $A_{true}$, as shown schematically in Fig. 3.

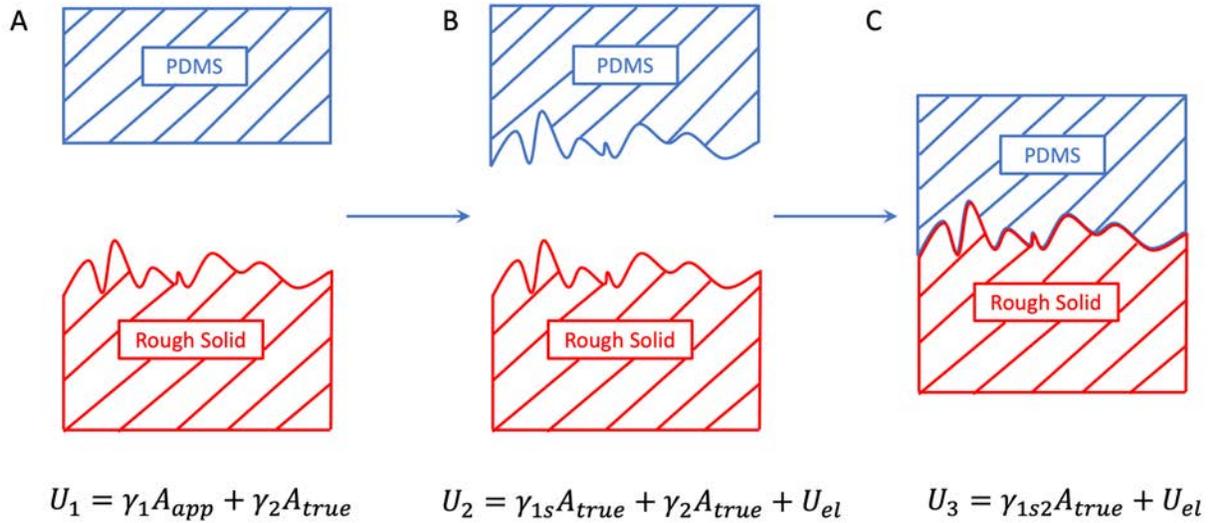

**Figure 3:** During adhesion, the materials go from the initial state (left) to the final state (right). However, to fully account for the energy change, one must consider the change in area of the soft material, which is represented schematically by including the intermediate state (middle).

To go from the initial state (Fig 3A) to the intermediate state (Fig. 3B), there is an energy change from $U_1$ to $U_2$. The PDMS is stretched and its surface energy changes depending upon the applied strain, which can be represented as a function of the area:

$$\Delta U_{1 \to 2} = \int_{A_{app}}^{A_{true}} \gamma_1(A) dA + U_{el} \tag{3}$$

Then, to go from the intermediate state $U_2$ (Fig. 3B) to the final state $U_3$ (Fig. 3C), there is an energy change of:

$$\Delta U_{2\rightarrow 3} = (\gamma_{1s2} - \gamma_{1s} - \gamma_2)A_{true} \quad (4)$$

Thus, the total work to go from the initial state to the final state is equal to $\Delta U_{1\rightarrow 2} + \Delta U_{2\rightarrow 3}$:

$$\Delta U_{1\rightarrow 3} = \int_{A_{app}}^{A_{true}} \gamma_1(A)dA + U_{el} + (\gamma_{1s2} - \gamma_{1s} - \gamma_2)A_{true} \quad (5)$$

This is the total energy change equal to $-A_{app}W_{app}$. Finally, we can re-write the total energy balance as:

$$W_{app}A_{app} = W_{int}^* A_{true} - \int_{A_{app}}^{A_{true}} \gamma_{1s}(A)dA - U_{el} \quad (6)$$

where, $W_{int}^* = \gamma_{1s} + \gamma_2 - \gamma_{1s2}$, and $\gamma_{1s}$ is the surface energy of the stretched elastomer. If we now make the assumption that the surface energy of the soft elastomer is not a strong function of strain (44), then $W_{int}^* = W_{int}$ and we can simplify the energy balance, and rearrange it to explicitly show $W_{app}$ as a function of two roughness-dependent terms, $A_{true}/A_{app}$ and $U_{el}/A_{app}$:

$$W_{app} = W_{int}\frac{A_{true}}{A_{app}} - \gamma_1\left(\frac{A_{true}}{A_{app}} - 1\right) - \frac{U_{el}}{A_{app}} \quad (7)$$

The stored elastic strain energy can be calculated from the power spectral density using the approach of Persson and Tosatti (24):

$$\frac{U_{el}}{A_{app}} = \frac{E^*}{8\pi}\int_0^\infty q^2 C^{iso}(q)\,dq \quad (8)$$

where, $C^{iso}$ is the radial average of the two-dimensional power-spectral density. For calculating the power spectral density, we follow the conventions used in Ref. (30). $C^{iso}$ is calculated from the one-dimensional PSD (Fig. 1) as described in the Methods. Finally, we derived a closed-form expression for the roughness-dependent increase in surface area, which works for arbitrary values of root-mean-square surface slope $h'_{rms}$ (SI Appendix, Section 3):

$$\frac{A_{true}}{A_{app}} = 1 + \frac{\sqrt{\pi}}{2}h'_{rms}\exp\left(\frac{1}{h'^2_{rms}}\right)\text{erfc}\left(\frac{1}{h'_{rms}}\right) \quad (9)$$

with $h'_{rms}$ calculated from the PSD as $(h'_{rms})^2 = \frac{1}{2\pi}\int_0^\infty q^3 C^{iso}(q)\,dq$ (30). For generality, all integrals were performed over the entire range of size scales over which topography was measured; if the range of wavevectors is instead cut off at the contact size (c.a. 100 microns), the extracted results are identical (within 0.1%). Taken together, Eqs. 7-9 demonstrate the predicted dependence of $W_{app}$ on material properties $(E, \nu)$ and topography $C^{iso}$.

The model for $W_{app}$ (Eq. 7) is applied to the measured data as shown in Fig. 4A using $\gamma_1 = 25 \pm 5$ mJ/m² for PDMS. This value was chosen based on prior work (41, 45), which also showed that the surface energy of PDMS does not change significantly with molecular weight. Furthermore, in the present investigation, water contact-angle measurements were performed on all PDMS materials and yielded values in the range of 103° - 107°, further supporting that all PDMS materials used in this investigation have similar surface energy. In applying this model, the minimum physically reasonable value of $W_{app}$ is set to zero; predicted values below zero (for 10 MPa PDMS on NCD and MCD) imply that the surfaces will not perfectly conform. The best correlation between the experimentally measured work of adhesion and the predictions of Eq. (7) was obtained using the intrinsic work of adhesion of $37.0 \pm 3.7$ mJ/m² ($R^2 = 0.67$). The reasonable value of $R^2$ (0.67) and the low value of standard error (3.7 mJ/m²) suggest good agreement between the model and the experimental measurements. The scatter in the experimental values as compared to the model is attributed to spot-to-spot variations. The theory outlined above cannot capture these spot-to-spot variations because it assumes a thermodynamic limit, corresponding to contacts of infinite size. The finite size of the experimental contact means that it is subject to finite-size fluctuations, such as a non-negligible probability for single anomalous asperities to dominate the response at low loads (46); this does not happen for theoretical contacts which sample the whole statistical distribution of the surface's roughness. Overall, the proposed model which explicitly accounts for the change in area of the soft surface (Eq. 7) achieves significantly improved model predictions; if we do not account for this change (calculations shown in SI Appendix, Section 4), the best fit to the measured data is significantly poorer ($R^2 = 0.29$).

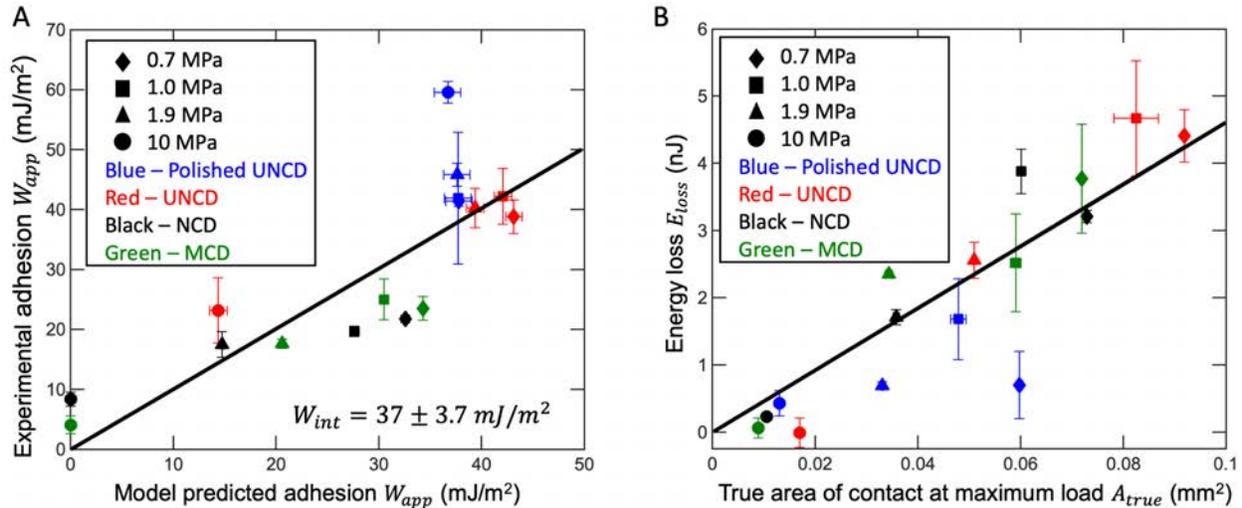

**Figure 4:** Comparison of work of adhesion and energy loss with the proposed model of conformal contact. In panel A, experimental measurements of apparent work of adhesion during approach are well-fit using the balance of adhesive and elastic energy described in the main text (Eqs. 7-9); here the solid line represents y = x. In panel B, the energy loss is plotted as a function of true contact area (Eq. 10). The solid line is a linear fit to the data and has a slope of $46.1 \pm 7.7$ mJ/m² ($R^2 = 0.8$).

The retraction portion of contact differs sharply from approach (as shown in Fig. 2), and the JKR model does not provide an adequate fit to the unloading data. Despite the poor fit, the JKR model can be used to extract a value for work of adhesion upon retraction, either by applying it only to the pull-off point, or by applying it to the several (c.a. 6) points before pull-off. Doing so (SI Appendix, Section 2) yields work of adhesion values in the range of 20 – 160 mJ/m². However, there is little consistency between these values and there is no connection to the intrinsic value of work of adhesion determined from the approach data.

Instead, we analyze the total energy loss during contact and separation. This quantity is computed as the integral under the loading and unloading curve, as shown in the inset in Fig. 2B. The *in situ* measurements of contact size yield the *apparent* area of contact during testing; to determine the true area of contact, we must multiply by the roughness-induced increase in true surface area (Eq. 9). We now plot the energy loss $E_{loss}$ versus the true area of contact $A_{true}$ *at maximum preload*. Figure 4B shows a linear correlation:

$$E_{loss} = W_{int} A_{true} \qquad (10)$$

with a best-fit intrinsic work of adhesion of 46.2 ± 7.7 mJ/m². The prior value of the work of adhesion (37.0 ± 3.7 mJ/m²) was measured during approach from the measured contact radius as a function of applied load. The latter value of work of adhesion (46.2 ± 7.7 mJ/m²) was obtained for the whole contact cycle (approach and separation) using the closed-circuit integral of the force-displacement curve. These values agree within their experimental uncertainties, despite being measured using very different approaches. This agreement suggests that we are indeed measuring an intrinsic work of adhesion for the materials, governed by the fundamental molecular interactions, rather than an effective property that may be governed by experimental parameters.

The present description of soft-material contact on rough surfaces assumes fully conformal contact. No method exists at present to directly verify this assumption; neither the *in situ* optical microscopy used in this investigation, nor the fluorescence or other techniques for imaging contact used elsewhere (e.g., Ref. (47)). However, the present results demonstrate that the experimentally measured behavior of these sixteen material pairs, during both loading and separation, is consistent with a model of conformal contact. This provides indirect evidence for the accuracy of this description of contact, and its underlying assumptions.

These results in Fig. 4 provide a simple physical mechanism to explain both the lower work of adhesion during approach and the adhesion hysteresis upon retraction. During approach, the apparent work of adhesion is reduced from $W_{int}$ by the energy required to deform the soft material to achieve conformal contact. This reduction can be quantitatively calculated using comprehensive, multi-scale measurements of topography (Eqs. 7-9). Furthermore, the energy loss during contact and separation matches with the product of $W_{int}$ and the *true* contact area $A_{true}$ at the maximum preload. Surface heterogeneities are known to pin the contact edge such that the retraction process depins the surface in instantaneous jumps over small localized microscopic regions (48). We show that Griffith's argument can be applied: these jumps occur once the elastic energy available is equal to the interface energy, and all elastic energy is dissipated in the creation of new surface (49).

Overall, the results show significant adhesion hysteresis in the absence of viscoelastic dissipation, and therefore demonstrate a fundamental origin of irreversible energy loss in soft materials that arises due to the roughness-induced increase in surface area and Griffith-like separation of the contact. This insight can be leveraged in applications involving reversible adhesion to real-world materials that contain roughness, such as the fields of soft robotics, biomaterials, and pick-and-place manufacturing. Equations 7-10 quantify the relative contributions to measured adhesion from material properties (intrinsic work of adhesion, elastic modulus, and Poisson ratio) and from surface topography (as characterized by a multi-scale PSD including atomic-scale information). This understanding suggests strategies to predict adhesion and to rationally modify it by tailoring the surface composition and surface topography.

# Methods

**Synthesis of Nanodiamond Substrates**
Nanodiamond films (Advanced Diamond Technologies, Romeoville, IL) were deposited using a tungsten hot-filament chemical vapor deposition (HF-CVD) system with parameters as described in Ref. (50). An H-rich gas mixture was used, with the chamber pressure of 5 Torr and a substrate temperature of 750 °C. The ratio of boron to carbon was maintained at 0.3 wt.%, to achieve high conductivity in the final film. The $CH_4$-to-$H_2$ ratio was modified (as described in Ref. (51)) to tune the grain size: achieving micro-, nano-, and ultrananocrystalline diamond. All films were grown to a thickness of 2 microns. Chemical-mechanical planarization was performed on an undoped UNCD film to create the polished UNCD samples. Previous surface analysis of synthetically grown diamond surfaces have shown similar surface composition regardless of grain size (52–54).

**Synthesis of PDMS hemispheres**
The smooth, soft elastic hemispheres were composed of cross-linked PDMS. To achieve systematic variation in modulus, we have used simple network theory, where changing the crosslinking molecular weight changes the crosslinking density and subsequently elastic modulus (34, 42), $E \sim \rho RT/M_c$, where $\rho$ is the density of the polymer, $R$ is the gas constant, $T$ is the temperature in Kelvin and $M_c$ is the cross-linked molecular weight. The curing system consisted of materials obtained from Gelest Inc.: vinyl-terminated PDMS of different molecular weights $M_w$ (DMS V-05 ($M_w$=800 gm/mol), V-21 ($M_w$=9000 gm/mol), V-31 ($M_w$=28000 gm/mol) and V-41 ($M_w$=62700 gm/mol)); tetrakis-dimethylsiloxysilane (SIT 7278.0) as tetra-functional cross-linker; platinum carbonyl cyclo-vinyl methyl siloxane complex (SIP 6829.2) as catalyst; 1,3,5,7-tetravinyl-1,3,5,7-tetramethyl cyclo-tetra siloxane (SIT 7900.0) as inhibitor. The vinyl-to-hydride molar ratio of 4.4 was maintained for all the samples avoiding excess cross-linker evaporation to minimize adhesion hysteresis from unreacted side chains as reported by Perutz *et al.* (43) The catalyst was added as 0.1% of the total batch. An additional reaction inhibitor was added to the DMS V-05 batch to avoid early cross-linking (5 times the catalyst amount). Hemispherical lenses were cast on the bottom of fluorinated glass dishes using a needle and a syringe. Since the PDMS mixture has a higher surface energy than the fluorinated surface, the drops maintain a contact angle on the surface giving a shape of a hemispherical lens. These lenses were imaged in profile using an optical microscope, and could be fit easily with a three-point circle to extract the necessary radius. They were cured at 60 °C for 3 days in a heating oven and Soxhlet-extracted using toluene at 124 °C for 24 hours. After 12 hours of drying in open air, the hemispheres were dried under vacuum at 120 °C overnight. The sol fraction for all of the batches was found to be less than 5%.

The fluorinated dishes were prepared by growing a monolayer of heptadecafluoro 1,1,2,2 tetrahydrodecatrichloro silane on clean base-bath-treated borosilicate glass petri-dishes. The octadecyltrichlorosilane (OTS) monolayer was prepared on silicon wafers (obtained from Silicon

Inc.), that had been pre-treated with piranha solution (3:7 ratio of 30% hydrogen peroxide: sulfuric acid (concentrated)). Silicon wafers were cleaned with an ample amount of water before use. The wafers were blown dry with nitrogen and plasma-treated before dipping in 1 wt.% OTS solution in toluene under nitrogen purge for 8 hours. The static water contact angle obtained was $110° ± 2°$ with negligible contact-angle hysteresis.

**Multi-scale Characterization of Surface Topography**
*Large-scale topography characterization: Stylus profilometry*
The largest scales of topography were measured using one-dimensional line scans with a stylus profilometer (Alpha Step IQ, KLA Tencor, Milpitas, CA) with a 5-µm diamond tip. Data were collected at a scanning speed of 10 µm/s, with data points every 100 nm. A total of 8 measurements were taken on each substrate, with 2 measurements each at scan sizes of 0.5, 1, 2, 5 mm. These measurements were taken at random orientations of the sample and did not show meaningful variations with direction. A parabolic correction was applied to all measurements which removed the tilt of the sample and the bowing artifact from the stylus tool. In two sessions (for the UNCD and polished UNCD), the larger scan sizes exhibited consistent non-parabolic trends due to instrument artifacts. In these cases, this was corrected by performing reference scans on polished silicon wafers and subtracting the averaged profiles from the reference measurements. Representative scans of stylus profilometry for all four materials are shown in SI Appendix, Fig. S1.

*Mid-scale topography characterization: Atomic force microscopy*
The substrates were measured using an atomic force microscope (Dimension V, Bruker, Billerica, MA) in tapping mode with diamond-like carbon-coated probes (Tap DLC300, Mikromasch, Watsonville, CA). For all substrates, square measurements were taken with the following lateral sizes: 3 scans each at 100 nm, 500 nm, and 5 µm; 1 scan each at 250 nm and 1 µm. The scanning speed was maintained at 1 µm/s for all scans. Each scan had 512 lines, with 512 data points per line, corresponding to pixel sizes in the range of 0.2 to 98 nm. The values of free-air amplitude and amplitude ratio were kept in the range of 37 – 49 nm and 0.15 – 0.3, respectively. While AFM provides a two-dimensional map of surface topography, the data were analyzed as a series of line scans, both to facilitate direct comparison with other techniques and to avoid apparent anisotropy due to instrument drift. Representative scans of atomic force microscopy for all four materials are shown in the left inset of Fig. 1.

*Small-scale topography characterization: Transmission electron microscopy*
Topography was measured on scales from microns to Ångströms following the approach developed in Ref. (55). For the UNCD, NCD, and MCD, the "wedge deposition technique" was used, whereas for polished UNCD, the "surface-preserving cross-section technique" was used (55). Briefly, the wedge deposition technique involves depositing the film of interest, in the same batch, on both flat silicon wafers (used for adhesion testing) and on standardized TEM-transparent silicon wedge samples (for TEM imaging). The surface-preserving cross-section technique is similar to conventional techniques for extraction of a TEM cross-section from a bulk sample (using grinding, polishing, dimple-grinding, and ion etching); however, modifications (especially to the ion etching step) ensure that the original surface topography is unmodified from its original state. The samples were imaged using a TEM (JEOL JEM 2100F, Tokyo, Japan) operated at 200 keV.

The images were taken with a 2000x2000-pixel camera using magnification levels from 5000x to 600,000x.

The nanoscale surface contours were extracted from the TEM images using custom Matlab scripts that create a digitized line profile based on a series of points selected by the user. The TEM images obtained were first rotated to make the surface horizontal and then the outer-most boundary was traced. While the vast majority of the measured surfaces were well-behaved functions *(i.e.,* there was a single value of height (y-axis) for each horizontal position (x-axis), there were some cases where two adjacent points were captured with identical or decreasing horizontal position. In these cases, the latter point was removed. In just 12 out of the 210 measurements, there were small portions of the profile that were reentrant. This character is not necessarily physically meaningful as it depends on the rotation of the TEM image during image analysis. Because the mathematical analyses (especially the calculation of PSD) require well-behaved functions, these regions were excluded from analysis.

**Combination of all measurements into complete, multi-scale PSD curves.**
For every topography measurement, the power spectral density was computed using the conventions described in Refs. (29, 30). The line scans from stylus profilometry, atomic force microscopy, and transmission electron microscopy all yield descriptions of the height h(x) over lateral position x. The Fourier transform of the surface topography is given by $\tilde{h}(q) = \int_0^L h(x)e^{-iqx}dx$; the PSD is computed as the square of the amplitude of $\tilde{h}(q)$; i.e., $C(q) = L^{-1}|\tilde{h}(q)|^2$. Since all collected data was analyzed as 1D line scans, then the computed PSDs were of the form of $C^{1D}$, using the nomenclature of Ref. (30).

For the tip-based measurement techniques (AFM and stylus), the tip-radius artifacts (30, 56) were eliminated using the criterion described in Eq. 2 of Ref. (29). This calculation was performed based on the tip radius measured using SEM images of the stylus tip and TEM images of the AFM probes after they were used for AFM measurement. Measured PSD data for wavevectors above the cutoff (calculated separately for each combination of tip and surface) were eliminated as unreliable. Finally, all reliable PSDs describing a single surface were combined into a single curve by computing the arithmetic average of the individually measured PSDs.

*Calculation of scalar roughness parameters*
The power spectral density can be integrated (as described in Refs. (29, 30)) to compute scalar descriptions of the surface: the root-mean-square height $h_{rms}$, RMS slope $h'_{rms}$, and RMS curvature $h''_{rms}$. The value of these parameters will depend on the scale over which they are measured (29); when all scales of topography are included, the computed values are as shown in SI Appendix, Table S1.

*Calculation of the two-dimensional PSD from the one-dimensional PSD*
All topography measurements in the present investigation are analyzed as 1D line-scans, and therefore the 1D PSD is presented in Fig. 1. However, the calculations proposed by Persson and Tosatti (and their modifications used in the present paper) employ a two-dimensional isotropic PSD. Under the assumption of isotropic roughness, the 2D PSD can be calculated from the 1D PSD, as described in Ref. (30). For this, we use Eq. A.28 of Ref. (30):

$$C^{iso}(q) \approx \frac{\pi}{q\sqrt{1-\left(\frac{q}{q_s}\right)^2}} C^{1D}(q) \qquad (11)$$

where $q_s$ is the short wavelength cut-off, in this case defined by the minimum wavelength at which roughness is measured (4 Å). This form of the 2D PSD is shown in SI Appendix, Fig. S3 and is used in the calculations for stored elastic energy and true surface area (Eqs. 8 and 9).

### *In situ* contact experiments and analysis
*Contact experiment methodology*
The contact experiment for each hemisphere-substrate combination was carried out using the setup shown in SI Appendix, Fig. S4 where simultaneous force and contact area measurement were taken during loading and unloading (Fig. 2). Optically transparent PDMS hemispheres of 2-3 mm diameter were used, with height greater than 700 μm to avoid substrate effects from the hemisphere's sample mount (57, 58). The maximum load applied for every measurement was 1 mN and the cycle was completed with a constant velocity of 60 nm/sec.

*Contact experiment analysis: Extracting values of work of adhesion*
To extract the apparent work of adhesion, the loading data is fit to the JKR equation (Eq. 1). These fits are shown as dashed lines in Fig. 2 and SI Appendix, Figs. S5 and S6. Since the contact radius, applied force, and radius of the lens $R$ are known, then the apparent work of adhesion and effective elastic modulus can be computed. The elastic moduli of the PDMS are calculated using the Poisson ratio for elastomers of 0.5, and using the modulus and Poisson ratio for diamond of 1100 GPa and 0.06, respectively (59). (However, because the effective modulus values are dominated by the properties of the elastomer, the precise values of diamond modulus are nearly insignificant.) The extracted values of modulus are those shown in the legend of Fig. 4, and are comparable to Ref. (34). The extracted values of $W_{app}$ are shown in SI Appendix, Table S2.

*Testing PDMS for adhesion hysteresis*
Before measuring work of adhesion on the rough nanodiamond substrates, the PDMS hemispheres were tested for inherent hysteresis against a low-surface energy OTS monolayer, which had been coated on a smooth silicon wafer. Plots of contact radius versus force are shown in SI Appendix, Fig. S6 for PDMS of different elastic moduli in contact with the OTS reference surface. The work of adhesion values obtained for loading and unloading fits are listed in SI Appendix, Table S3 showing comparable values and low hysteresis. This OTS reference testing was repeated before and after the measurements on the nanodiamond substrates to rule out any permanent changes in the cross-linked structure of PDMS due to testing. The hemispheres did not show significant deviation from the original numbers in the post-test measurements. The closed-circuit integral for the force-displacement curve (calculated as shown in the inset of Fig. 2B) for PDMS-OTS have values that are at least an order of magnitude smaller than those measured on the rough surfaces (SI Appendix, Table S3).

Data from this publication will be made publicly available upon acceptance of the paper.

## Acknowledgements

Dr. K. Vorovolakos, Prof. M. K. Chaudhury and Prof. C. Cohen provided necessary guidance to make soft materials. We thank Dr. B. N. J. Persson, Daniel Maksuta, Michael C. Wilson and Antoine Sanner for helpful comments. We thank Prof. K. Beschorner for help with statistical analysis. T.D.B.J. acknowledges support from the NSF under award number CMMI-1727378. A.D. acknowledges funding from NSF (DMR-1610483). Use of the Nanoscale Fabrication and Characterization Facility (NFCF) in the Petersen Institute for Nano Science and Engineering (PINSE) is acknowledged.

A.D. and T.D.B.J designed the experiments. S.D. synthesized soft materials, conducted and analyzed the adhesion results. A.G. and S.R.K measured the surface roughness and computed the PSDs for the nanodiamond surfaces. A.G. and L.P. conducted the theoretical analysis of surface roughness and A.D. proposed the modified energy balance. All authors contributed in writing the paper.

# References

1. Ayyildiz M, Scaraggi M, Sirin O, Basdogan C, Persson BNJ (2018) Contact mechanics between the human finger and a touchscreen under electroadhesion. *Proc Natl Acad Sci U S A* 115(50):12668–12673.
2. Niewiarowski PH, Stark AY, Dhinojwala A (2016) Sticking to the story: Outstanding challenges in gecko-inspired adhesives. *J Exp Biol* 219(7):912–919.
3. Persson BNJ, Albohr O, Tartaglino U, Volokitin AI, Tosatti E (2005) On the nature of surface roughness with application to contact mechanics, sealing, rubber friction and adhesion. *J Phys Condens Matter* 17(1):R1.
4. Luan B, Robbins MO (2005) The breakdown of continuum models for mechanical contacts. *Nature* 435(7044):929–932.
5. Jacobs TDB, et al. (2013) The effect of atomic-scale roughness on the adhesion of nanoscale asperities: A combined simulation and experimental investigation. *Tribol Lett* 50(1):81–93.
6. Pastewka L, Robbins MO (2014) Contact between rough surfaces and a criterion for macroscopic adhesion. *Proc Natl Acad Sci* 111(9):1–6.
7. Popa DO, Stephanou HE (2004) Micro and mesoscale robotic assembly. *J Manuf Process* 6(1):52–71.
8. Carlson A, Bowen AM, Huang Y, Nuzzo RG, Rogers JA (2012) Transfer printing techniques for materials assembly and micro/nanodevice fabrication. *Adv Mater* 24(39):5284–5318.
9. Yu Z, Cheng H (2018) Tunable adhesion for bio-integrated devices. *Micromachines* 9(10):529.
10. Tang Y, Zhang Q, Lin G, Yin J (2018) Switchable Adhesion Actuator for Amphibious Climbing Soft Robot. *Soft Robot* 5(5):592–600.
11. Hertz H (1882) On the contact of rigid elastic solids. *J fur die Reine und Angew Math*:449.
12. Johnson KL, Kendall K, Roberts AD (1971) Surface Energy and the Contact of Elastic Solids. *Proc R Soc A Math Phys Eng Sci* 324(1558):301–313.
13. Israelachvili JN (2015) *Intermolecular and Surface Forces* (Academic Press (London)). 3rd Ed.

14. Briggs GAD, Briscoe BJ (1977) The effect of surface topography on the adhesion of elastic solids. *J Phys D Appl Phys* 10(18):2453–2466.
15. Fuller KNG, Tabor D (1975) The effect of surface roughness on the adhesion of elastic solids. *Proc R Soc London A Math Phys Sci* 345(1642):327–342.
16. Maeda N, Chen N, Tirrell M, Israelachvili JN (2002) Adhesion and friction mechanisms of polymer-on-polymer surfaces. *Science (80- )* 297(5580):379–382.
17. Luengo G, Pan J, Heuberger M, Israelachvili JN (1998) Temperature and Time Effects on the "Adhesion Dynamics" of Poly(butyl methacrylate) (PBMA) Surfaces. *Langmuir* 14(14):3873–3881.
18. Tiwari A, et al. (2017) The effect of surface roughness and viscoelasticity on rubber adhesion. *Soft Matter* 13:3602–3621.
19. Archard J. F. (1957) Elastic deformation and the laws of friction. *Proc R Soc A Math Phys Eng Sci* 243(1233):190–205.
20. Greenwood JA, Williamson JBP (1966) Contact of Nominally Flat Surfaces. *Proc R Soc London A Math Phys Eng Sci* 295(1442):300–319.
21. Jackson RL, Streator JL (2006) A multi-scale model for contact between rough surfaces. *Wear* 261(11–12):1337–1347.
22. Thomas TR, Sayles RS (1978) Some problems in the tribology of rough surfaces. *Tribol Int* 11(3):163–168.
23. Benoit B Mandelbrot, Dann E Passoja AJP (1991) Fractal Character of Fracture Surfaces of Metals. *Nature* 308(5961):721.
24. Persson BNJ, Tosatti E (2001) The effect of surface roughness on the adhesion of elastic solids. *J Chem Phys* 115(12):5597–5610.
25. Yang C, Persson BNJ (2008) Molecular dynamics study of contact mechanics: Contact area and interfacial separation from small to full contact. *Phys Rev Lett* 100(2):024303.
26. Persson BNJ, Scaraggi M (2014) Theory of adhesion: Role of surface roughness. *J Chem Phys* 141(12):124701.
27. Hyun S, Pei L, Molinari JF, Robbins MO (2004) Finite-element analysis of contact between elastic self-affine surfaces. *Phys Rev E - Stat Physics, Plasmas, Fluids, Relat Interdiscip Top* 70:026117.
28. Putignano C, Afferrante L, Carbone G, Demelio G (2012) The influence of the statistical properties of self-affine surfaces in elastic contacts: A numerical investigation. *J Mech Phys Solids* 60(5):973–982.
29. Gujrati A, Khanal SR, Pastewka L, Jacobs TDB (2018) Combining TEM, AFM, and Profilometry for Quantitative Topography Characterization Across All Scales. *ACS Appl Mater Interfaces* 10(34):29169–29178.
30. Jacobs TDB, Junge T, Pastewka L (2017) Quantitative characterization of surface topography using spectral analysis. *Surf Topogr Metrol Prop* 5:013001.
31. Zhang X, Xu Y, Jackson RL (2017) An analysis of generated fractal and measured rough surfaces in regards to their multi-scale structure and fractal dimension. *Tribol Int* 105:94–101.
32. Silberzan P, Perutz S, Kramer EJ, Chaudhury MK (1994) Study of the Self-Adhesion Hysteresis of a Siloxane Elastomer Using the JKR Method. *Langmuir* 10(7):2466–2470.
33. Choi GY, Kim S, Ulman A (1997) Adhesion Hysteresis Studies of Extracted Poly(dimethylsiloxane) Using Contact Mechanics. *Langmuir* 13(23):6333–6338.
34. Vorvolakos K, Chaudhury MK (2003) The effects of molecular weight and temperature

on the kinetic friction of silicone rubbers. *Langmuir* 19(17):6778–6787.
35. Chen YL, Helm CA, Israelachvili JN (1991) Molecular mechanisms associated with adhesion and contact angle hysteresis of monolayer surfaces. *J Phys Chem* 95(26):10736–10747.
36. Pickering JP, Van Der Meer DW, Vancso GJ (2001) Effects of contact time, humidity, and surface roughness on the adhesion hysteresis of polydimethylsiloxane. *J Adhes Sci Technol* 15(12):1429–1441.
37. Kesari H, Doll JC, Pruitt BL, Cai W, Lew AJ (2010) Role of surface roughness in hysteresis during adhesive elastic contact. *Philos Mag Lett* 90(12):891–902.
38. Peressadko AG, Hosoda N, Persson BNJ (2005) Influence of surface roughness on adhesion between elastic bodies. *Phys Rev Lett* 95(12):1–4.
39. Yurdumakan B, Harp GP, Tsige M, Dhinojwala A (2005) Template-induced enhanced ordering under confinement. *Langmuir* 21(23):10316–10319.
40. She H, Malotky D, Chaudhury MK (1998) Estimation of Adhesion Hysteresis at Polymer/Oxide Interfaces Using Rolling Contact Mechanics. *Langmuir* 14(11):3090–3100.
41. Chaudhury MK (1996) Interfacial interaction between low-energy surfaces. *Mater Sci Eng R Reports* 16(3):97–159.
42. Vaenkatesan V, Li Z, Vellinga WP, de Jeu WH (2006) Adhesion and friction behaviours of polydimethylsiloxane - A fresh perspective on JKR measurements. *Polymer (Guildf)* 47(25):8317–8325.
43. Perutz S, Kramer EJ, Baney J, Hui CY, Cohen C (1998) Investigation of adhesion hysteresis in poly(dimethylsiloxane) networks using the JKR technique. *J Polym Sci Part B Polym Phys* 36(12):2129–2139.
44. Liang H, Cao Z, Wang Z, Dobrynin A V. (2018) Surface Stress and Surface Tension in Polymeric Networks. *ACS Macro Lett* 7(1):116–121.
45. Chaudhury MK, Whitesides GM (1992) Correlation between surface free energy and surface constitution. *Science (80- )* 255(5049):1230–1232.
46. Pastewka L, Robbins MO (2016) Contact area of rough spheres: Large scale simulations and simple scaling laws. *Appl Phys Lett* 108(22):221601.
47. Weber B, et al. (2018) Molecular probes reveal deviations from Amontons' law in multi-asperity frictional contacts. *Nat Commun* 9(1):888.
48. Persson BNJ (2003) Nanoadhesion. *Wear* 254(9):832–834.
49. Griffith AA (1921) The Phenomena of Rupture and Flow in Solids. *Philos Trans R Soc A Math Phys Eng Sci* 221(582–593).
50. Zeng H, et al. (2015) Boron-doped ultrananocrystalline diamond synthesized with an H-rich/Ar-lean gas system. *Carbon N Y* 84:103–117.
51. Auciello O, et al. (2007) Are diamonds a MEMS' best friend? *IEEE Microw Mag* 8(6):61–75.
52. Sumant A V., et al. (2005) Toward the ultimate tribological interface: Surface chemistry and nanotribology of ultrananocrystalline diamond. *Adv Mater* 17(8):1039–1045.
53. Härtl A, et al. (2004) Protein-modified nanocrystalline diamond thin films for biosensor applications. *Nat Mater* 3(10):736–742.
54. Fuentes-Fernandez EMA, et al. (2016) Synthesis and characterization of microcrystalline diamond to ultrananocrystalline diamond films via Hot Filament Chemical Vapor Deposition for scaling to large area applications. *Thin Solid Films* 603:62–68.

55. Khanal SR, et al. (2018) Characterization of small-scale surface topography using transmission electron microscopy. *Surf Topogr Metrol Prop* 6:045004.
56. Church EL, Takacs PZ (1991) Effects of the nonvanishing tip size in mechanical profile measurements. *Proc SPIE 1332, Opt Test Metrol III Recent Adv Ind Opt Insp* 1332:504–514.
57. Deruelle M, Hervet H, Jandeau G, Léger L (1998) Some remarks on JKR experiments. *J Adhes Sci Technol* 12:225–247.
58. Wald MJ, Considine JM, Turner KT (2013) Determining the Elastic Modulus of Compliant Thin Films Supported on Substrates from Flat Punch Indentation Measurements. *Exp Mech* 53(6):931–941.
59. Klein CA, Cardinale GF (1993) Young's modulus and Poisson's ratio of CVD diamond. *Diam Relat Mater* 2(5–7):918–923.

# Supplemental Information Appendix for "Linking energy loss in soft adhesion to surface roughness"

Siddhesh Dalvi[1*], Abhijeet Gujrati[2*], Subarna R. Khanal[2], Lars Pastewka[3], Ali Dhinojwala[1#], Tevis D. B. Jacobs[2#]

[1]Department of Polymer Science, The University of Akron, Akron, Ohio, 44325, United States
[2]Department of Mechanical Engineering and Materials Science, University of Pittsburgh, Pittsburgh, Pennsylvania, 15261, United States
[3]Department of Microsystems Engineering, University of Freiburg, 79110 Freiburg, Germany

*these two authors contributed equally
#corresponding authors

## SI Appendix, Section 1. Additional details on the characterization of surface topography of the nanodiamond substrates.

Examples of stylus profilometry are shown in Fig. S1. Additional power spectral density data are included in Figs. S2-3 to demonstrate exact regions of applicability of each technique (Fig. S2) and to show the 2-D power spectral density (Fig. S3), which is calculated from the 1D PSD as described in the Methods.

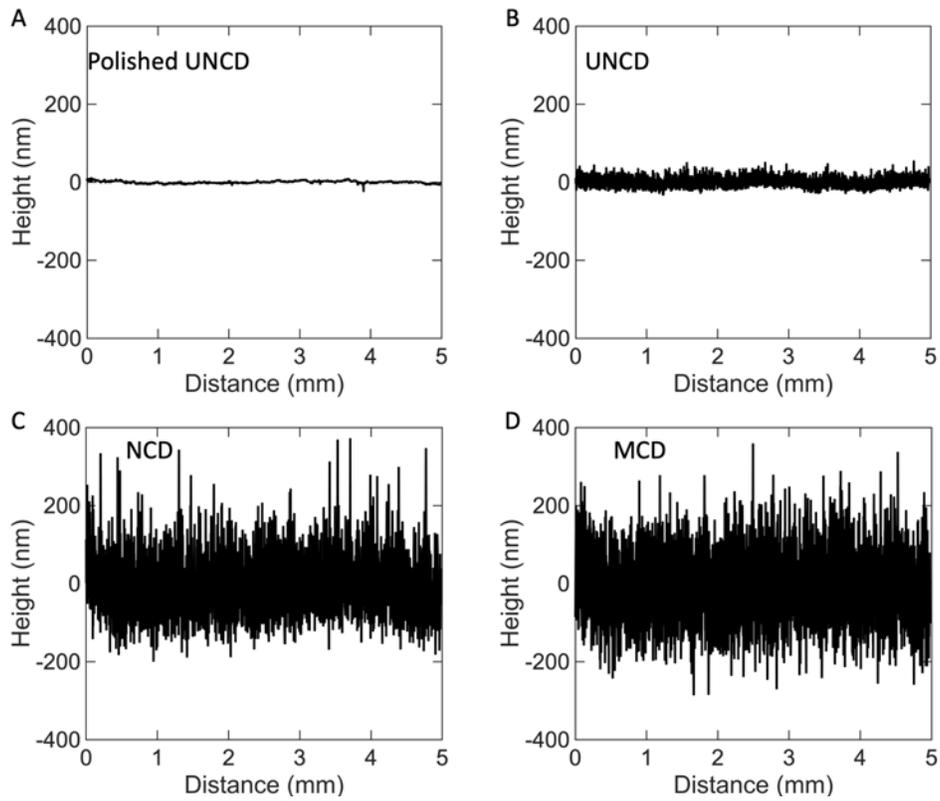

**Figure S1:** Stylus profilometry of the polished UNCD (A), UNCD (B), NCD (C) and MCD (D).

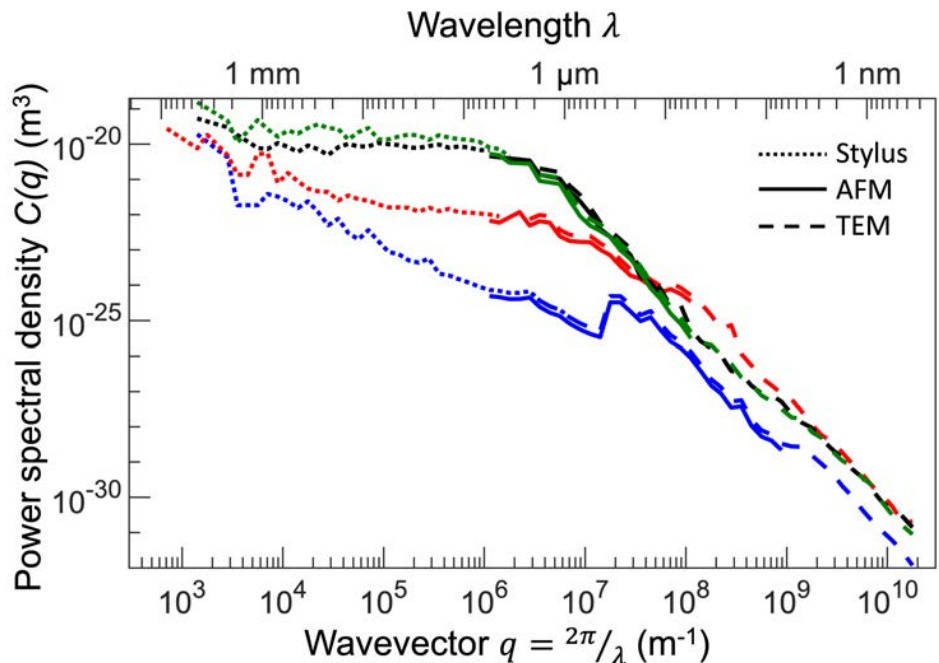

**Figure S2:** Power spectral densities of the four surfaces, with indication of the specific regimes of applicability of each technique. The curves on this plot represent the identical data to Fig. 1 of the main text, however the present figure uses line style (solid, dashed, dotted) to indicate the specific bandwidth over which different techniques were applied. Because of the nature of tip artifacts, the minimum size from stylus and AFM data differ between surfaces.

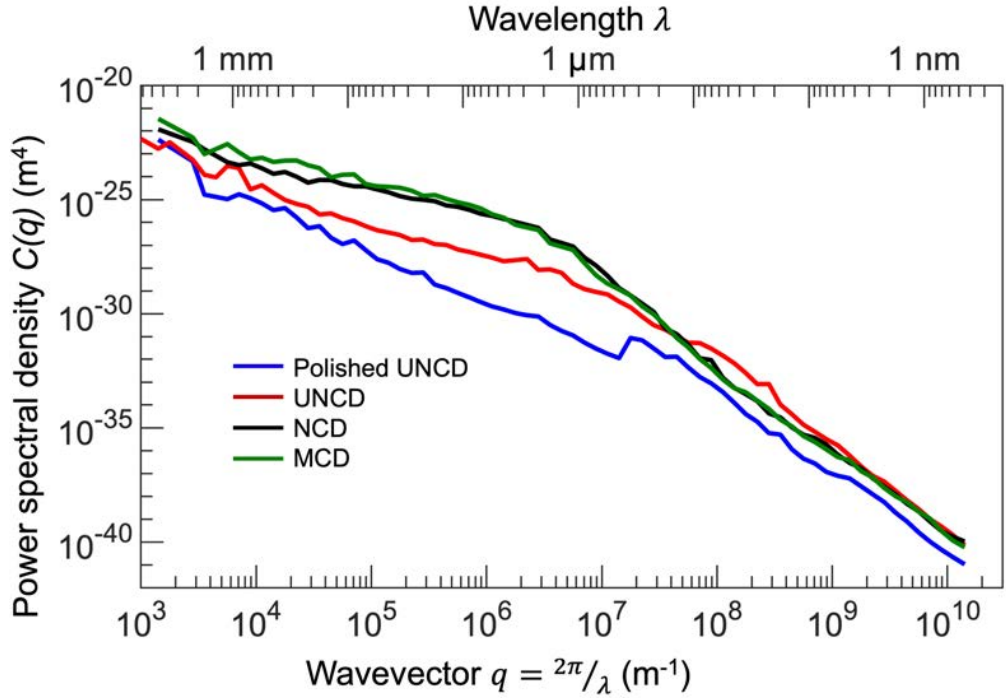

**Figure S3:** 2D power spectral densities, after conversion from the 1D values shown in Fig. 1 of the main text.

The RMS parameters for each surface were computed in frequency space from integrals over the entire PSD (1). This has the advantage over computing them in real-space because any real-space measurement is taken with only a finite bandwidth, and the absolute value of the parameters varies by orders of magnitude with the size scale over which they are measured (2). The RMS parameters can be computed in 1D (using $C^{1D}$ and Eq. 1 of Ref. (2)) or in 2D (using $C^{iso}$ and Eqs. 4, 8 and 9 of Ref. (1)). Table S1 reports the 2D RMS parameters.

**Table S1:** RMS parameters (2D) for nanodiamond substrates.

|  | Polished UNCD | UNCD | NCD | MCD |
| --- | --- | --- | --- | --- |
| RMS height | 4.6 ± 0.8 nm | 23.4 ± 1.3 nm | 121.7 ± 13.4 nm | 126.6 ± 8.2 nm |
| RMS slope | 0.39 ± 0.04 | 1.46 ± 0.36 | 1.15 ± 0.13 | 1.07 ± 0.13 |
| RMS curvature | 1.13 ± 0.23 nm$^{-1}$ | 3.37 ± 0.69 nm$^{-1}$ | 3.19 ± 1.15 nm$^{-1}$ | 2.83 ± 0.81 nm$^{-1}$ |

# SI Appendix, Section 2. *In situ* contact experiment and analysis

The experimental test setup is shown schematically in Fig. S4. Contact radius results from all materials are shown in Fig. S5

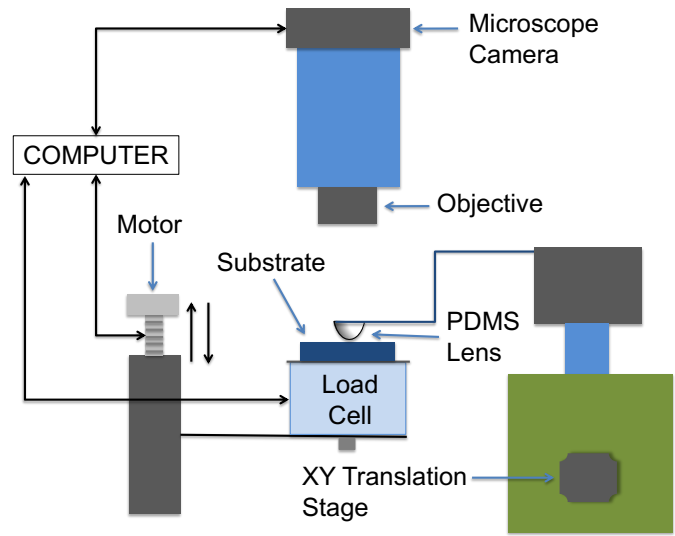

**Figure S4:** Schematic of the *in situ* apparatus used to measure work of adhesion and elastic modulus

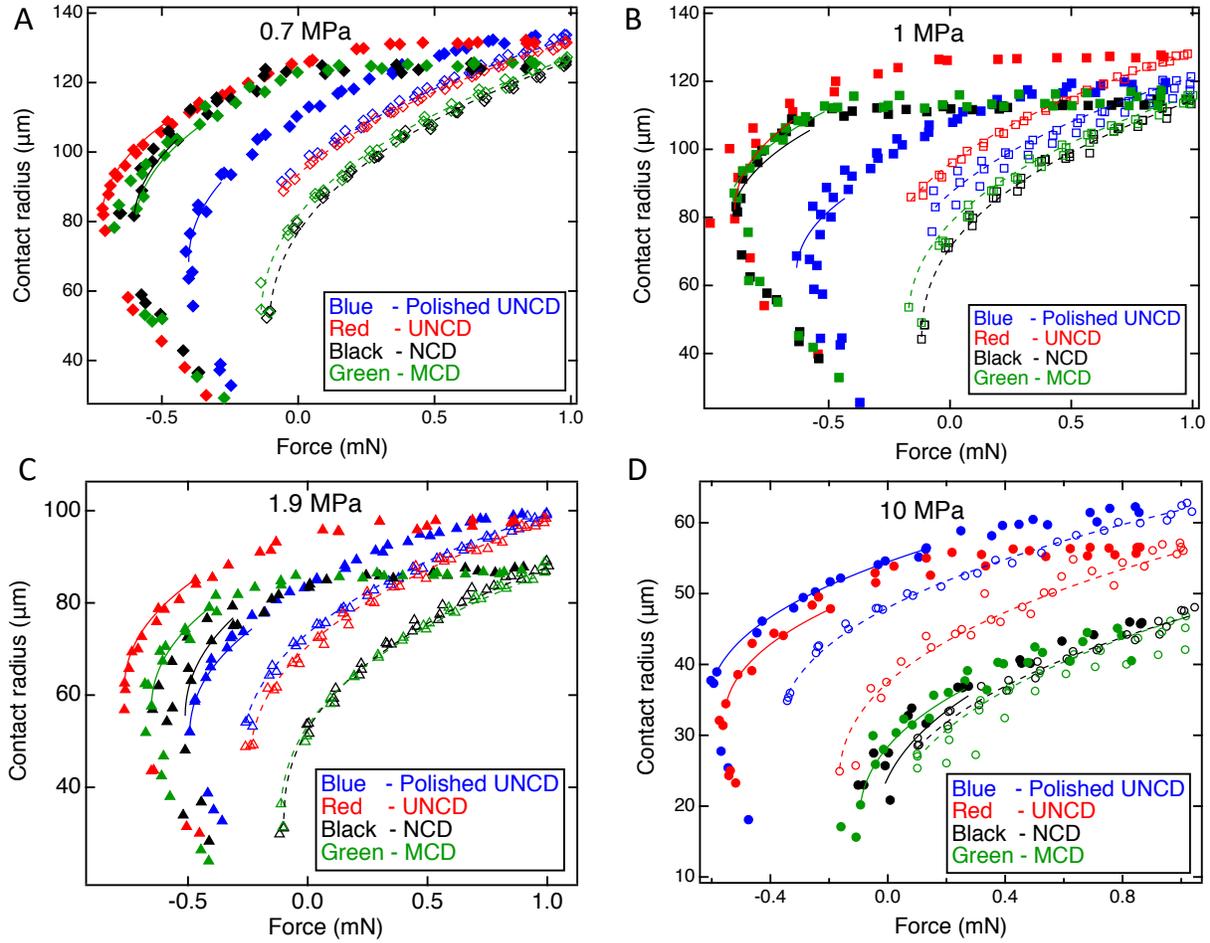

**Figure S5:** Contact radius was measured as a function of applied force plots for PDMS spheres with elastic modulus of 0.7 MPa (A), 1 MPa (B), 1.9 MPa (C), and 10 MPa (D). The loading data are represented using hollow symbols and are fit using Eq. 1 (dashed line) to extract the apparent work of adhesion $W_{app}$. The separation data are represented using filled symbols; a subset of the data are fit using Eq. S1 (solid line) to extract $W_{app,ret}$.

As mentioned in the main text, the retraction portions of the experiments on the nanodiamond substrate *do not* follow the trends of the JKR model. However, values of the work of adhesion on retraction $W_{app,ret}$ can be extracted by force-fitting the JKR model to the data. This can be done in one of two ways. First, the work of adhesion can be calculated using the simple JKR equation for the theoretical value of maximum pull-off force $F_{po}$, which is(3):

$$F_{po} = -\frac{3}{2}\pi R W_{\text{app,ret}} \qquad (S1)$$

Alternatively, the JKR equation can be rearranged to be a function of $F_{po}$, and this equation can be fit to the both approach and retraction data(4):

$$a = \left[\frac{3RF_{po}}{4E^*}\left(\frac{F}{F_{po}} + 2 + 2\left\{\frac{F}{F_{po}} + 1\right\}^{1/2}\right)\right]^{1/3} \qquad (S2)$$

While this function cannot be fit to the entire unloading portion, it can be fit to several points near the point of pull-off. Both of these two approaches (Eqs. S1 and S2) yield similar values for $W_{app,ret}$, as shown in Table S2.

**Table S2:** Comparison of different work of adhesion values for nanodiamond substrates.

| Substrate | Work of Adhesion upon Approach (in mJ/m$^2$) | | | |
|---|---|---|---|---|
| | E = 0.69 MPa ± 0.02 MPa | E = 1.03 MPa ± 0.02 MPa | E = 1.91 MPa ± 0.11 MPa | E = 10.03 MPa ± 0.88 MPa |
| PUNCD | 41.41 ± 0.86 | 41.91 ± 10.97 | 45.82 ± 1.92 | 59.55 ± 1.82 |
| UNCD | 38.82 ± 2.81 | 42.22 ± 4.66 | 40.28 ± 3.28 | 23.15 ± 5.46 |
| NCD | 21.73 ± 0.60 | 19.64 ± 0.86 | 17.47 ± 2.15 | 8.37 ± 1.12 |
| MCD | 23.49 ± 1.97 | 24.98 ± 3.39 | 17.60 ± 0.79 | 4.06 ± 1.46 |
| Work of Adhesion from Pull-off using eq. S3 (in mJ/m$^2$) | | | | |
| PUNCD | 74.73 ± 2.58 | 87.97 ± 2.31 | 83.04 ± 2.00 | 102.01 ± 0.96 |
| UNCD | 153.03 ± 2.46 | 147.60 ± 17.10 | 131.65 ± 0.98 | 94.40 ± 1.29 |
| NCD | 118.26 ± 5.07 | 142.02 ± 5.76 | 100.87 ± 10.65 | 17.18 ± 4.74 |
| MCD | 120.03 ± 8.19 | 144.95 ± 7.26 | 116.01 ± 3.43 | 21.38 ± 4.99 |
| Work of Adhesion upon Retraction using eq. S4 (in mJ/m$^2$) | | | | |
| PUNCD | 72.67 ± 2.2 | 95.2 ± 6.8 | 80.65 ± 1.72 | 94.4 ± 1.35 |
| UNCD | 131.67 ± 1.7 | 143.93 ± 16.3 | 128.43 ± 0.06 | 88.21 ± 1.18 |
| NCD | 116.2 ± 5.76 | 144.01 ± 2.27 | 97.82 ± 11.5 | 13.76 ± 5.7 |
| MCD | 118.74 ± 8.7 | 142.45 ± 7.38 | 113.39 ± 3.71 | 19.0 ± 5.22 |

*Testing PDMS for adhesion hysteresis*

As mentioned in the main text, to test for adhesion hysteresis due to material viscoelasticity, the PDMS spheres were tested against a smooth silicon wafer coated with a low-surface energy octadecyltrichlorosilane (OTS) monolayer. The results are shown in Fig. S6 and quantified in Table S3.

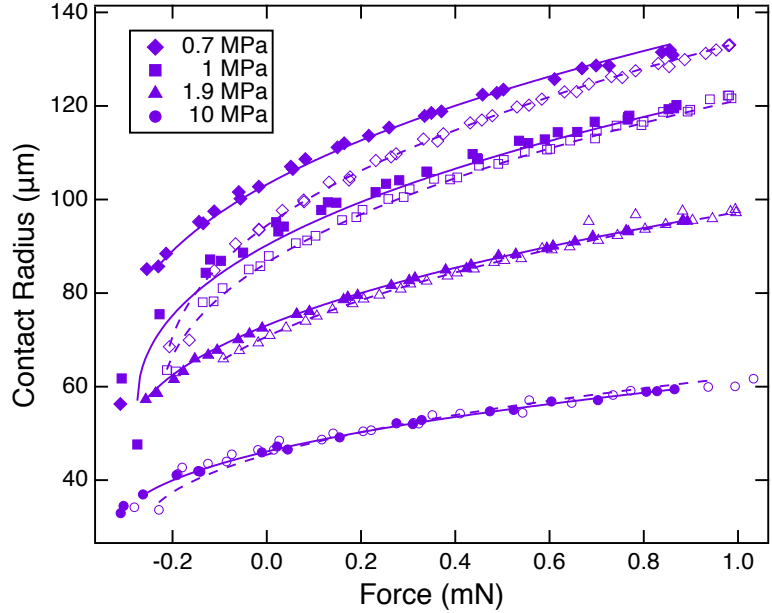

**Figure S6:** The contact radius data for the PDMS hemispheres on the OTS surface show low hysteresis between loading (empty symbols) and unloading (filled symbols). The dashed lines indicate JKR model fits for loading and solid lines indicate the JKR model fits for unloading.

**Table S3:** Work of adhesion and excess energy measurements for the OTS reference substrate.

| $M_c$ (gm/mol) | $W_{PDMS-OTS}$ (approach) (mJ/m²) | Elastic Modulus (MPa) | $W_{PDMS-OTS}$ (retraction) (mJ/m²) | Excess Energy (nJ) |
|---|---|---|---|---|
| 800 | 51.0 ± 4.8 | 10.0 ± 0.9 | 56.4 ± 1.8 | 1.59 ± 0.91 |
| 6000 | 38.8 ± 2.9 | 1.9 ± 0.1 | 52.2 ± 1.3 | 0.45 ± 0.45 |
| 28000 | 36.8 ± 0.8 | 1.0 ± 0.0 | 52.5 ± 4.8 | 0.49 ± 0.29 |
| 62700 | 39.6 ± 1.2 | 0.7 ± 0.0 | 59.3 ± 1.0 | 0.38 ± 0.02 |

## SI Appendix, Section 3. Deriving an expression for the increase in surface area due to roughness for large slopes.

Prior work (e.g. Ref. (5)) has derived expressions for $A_{\text{true}}/A_{\text{app}}$ in the limit of small slopes. Here, we derive an expression for $A_{\text{true}}/A_{\text{app}}$ that works for arbitrary values of slope $h'_{\text{rms}}$. The derivation follows along the arguments given in the Supplementary Material of Ref. (2).

For a full two-dimensional topography map $h(x, y)$, the surface area $A_{\text{true}}$ is straightforwardly obtained from an expression analogous to the arc length of a function:

$$A_{\text{true}} = \int_{A_{\text{app}}} \sqrt{1 + |\nabla h|^2}\, dxdy = \int_{A_{\text{app}}} \sqrt{1 + \left(\frac{\partial h}{\partial x}\right)^2 + \left(\frac{\partial h}{\partial y}\right)^2}\, dxdy \qquad (S3)$$

For small slopes $|\nabla h|$, the square-root can be expanded into a Taylor series and truncated above quadratic order. This gives the well-known expression (5):

$$A_{\text{true}} \approx \int_{A_{\text{app}}} \left(1 + \frac{1}{2}|\nabla h|^2\right) dx\,dy = A_{\text{app}}\left(1 + \frac{1}{2}h'^2_{\text{rms}}\right) \tag{S4}$$

with

$$h'^2_{\text{rms}} = \int_{A_{\text{app}}} |\nabla h|^2 dx\,dy \tag{S5}$$

In order to arrive at an expression valid for large $h'_{\text{rms}}$, we now transform the integral over the surface area into an integral over slopes. We first define the slope distribution function,

$$\phi(s_x, s_y) = \frac{1}{A_{\text{app}}} \int_{A_{\text{app}}} \delta\left(s_x - \frac{\partial h(x,y)}{\partial x}\right) \delta\left(s_y - \frac{\partial h(x,y)}{\partial y}\right) dx\,dy, \tag{S6}$$

where $\delta(x)$ is the Dirac delta function. Note that using the slope distribution function, we can express the integral over any function $f$ that depends on just slopes as

$$\int_{A_{\text{app}}} f\left(\frac{\partial h}{\partial x}, \frac{\partial h}{\partial y}\right) dx\,dy = A_{\text{app}} \int \phi(s_x, s_y) f(s_x, s_y) ds_x\,ds_y \tag{S7}$$

We can hence re-express Eq. S3 as:

$$\frac{A_{\text{true}}}{A_{\text{app}}} = \int \phi(s_x, s_y) \sqrt{1 + s_x^2 + s_y^2}\, ds_x\,ds_y \tag{S8}$$

We now make the assumption that our surfaces are isotropic and Gaussian. The slope distribution function is then given by

$$\phi(s_x, s_y) = \frac{1}{\pi h'^2_{\text{rms}}} \exp\left(-\frac{s_x^2 + s_y^2}{h'^2_{\text{rms}}}\right) \tag{S9}$$

with (see also Eq. S5)

$$h'^2_{\text{rms}} = \int_{A_{\text{app}}} |\nabla h|^2 dx\,dy = \int \phi(s_x, s_y)(s_x^2 + s_y^2) ds_x\,ds_y. \tag{S10}$$

Evaluating Eq. S6 using this slope distribution function yields

$$\frac{A_{\text{true}}}{A_{\text{app}}} = \frac{2}{h'^2_{\text{rms}}} \int_0^\infty \exp\left(-\frac{s^2}{h'^2_{\text{rms}}}\right) \sqrt{1+s^2}\, s\,ds = 1 + \frac{1}{2} h'^2_{\text{rms}} g(h'_{\text{rms}}) \tag{S11}$$

with

$$g(h'_{\text{rms}}) = \sqrt{\pi} \exp\left(\frac{1}{h'^2_{\text{rms}}}\right) \text{erfc}\left(\frac{1}{h'_{\text{rms}}}\right)/h'_{\text{rms}} \tag{S12}$$

Equation S11 is Eq. 9 from the main text. Note that the left-hand side of Eq. S11 is Eq. B1 from Ref. (5). The function $g(h'_{\text{rms}})$ can be regarded a correction to the small slope approximation Eq. S4. It has the property $g(h'_{\text{rms}}) \to 1$ as $h'_{\text{rms}} \to 0$ and hence we recover Eq. S4 from Eq. S11 in the small slope limit. Figure S7A shows the function $g$ up to slope of 5. Note that for slope of order unity, $g(1) \approx 0.76$ and hence the small-slope approximation Eq. S4 would overestimate the area by 30%.

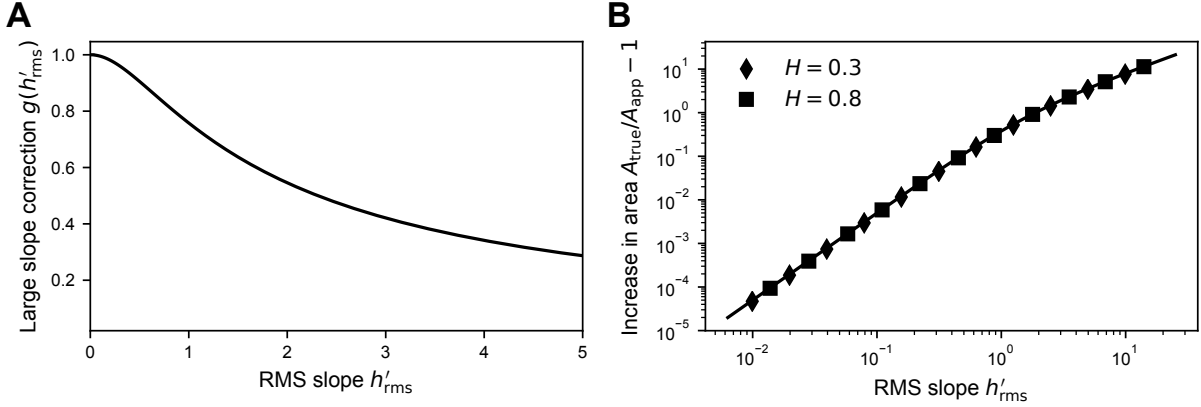

**Figure S7:** Plot of the correction $g(h'_{rms})$ to the small-slope approximation. For values of $g(h'_{rms}) \approx 1$ the small slope approximation is valid (A) Validation of Eq. S11 using computer-generated self-affine surfaces with varying RMS slope $h'_{rms}$ and Hurst exponents $H$ (B) The solid line shows the analytic result given by Eq. S11.

In order to numerically test the validity of Eq. S11, we have created a range of synthetic self-affine surfaces with 4096 x 4096 points and Hurst exponent $H = 0.3$ and 0.8 using a Fourier filtering algorithm (1, 6). We then computed the true surface area by numerical integration of Eq. S4. Figure S7B shows that the analytic expression Eq. S11 describes the synthetic surfaces excellently up to slopes of order 10.

## SI Appendix, Section 4. Calculating work of adhesion without accounting for the change in area of the soft material

The original model by Persson and Tosatti (5) leads to an equation for $W_{app}$ in terms of topography:

$$W_{\text{app}} = W_{\text{int}} \left\{ \left[ 1 + \frac{1}{4\pi} \int_{q_0}^{q_1} dq\, q^3 C^{iso}(q) \right] - \left[ \frac{E}{8\pi(1-\nu^2)W_{int}} \int_{q_0}^{q_1} dq\, q^2 C^{iso}(q) \right] \right\} \quad (S13)$$

where $q_0$ is the long-wavelength (small-wavevector) cut-off and $q_1$ is the short-wavelength (large-wavevector) cut-off of the topography. Note that, as described in Ref. (1), $C^{iso}$ differs by a constant prefactor from the PSD definition used by Persson and Tosatti, arising from different conventions used in the Fourier transform. Therefore, the prefactors in Eq. S3 differ from those in Ref. (5). These differences can be reconciled by acknowledging (1) that $C^{Persson}(q) = C^{iso}(q)/4\pi^2$. In Ref. (5), Eq. S3 is further simplified for self-affine surfaces. However, in this study we have directly used the integral equations because the surfaces are not self-affine over all length-scales.

To calculate $W_{app}$ using the combined PSD from the model, Eq. S13 was integrated using the data in Fig. S3. The wavevector cutoffs were set as the maximum and minimum measured values ($q_0 = 1.3 \times 10^3$ m$^{-1}$ and $q_1 = 1.6 \times 10^{10}$ m$^{-1}$). Figure S8 shows the experimentally measured values of $W_{app}$ compared against the predictions of Eq. S13. The best fit was obtained using $W_{int} = 25$ mJ/m$^2$. The proposed model (main text) is considered to more accurately describe the present data as compared to the Persson-Tosatti model for three reasons: first, it more accurately accounts for the change in area of the PDMS; second, the fit to the data is better ($R^2 = 0.29$ for the Persson-Tosatti model and $R^2 = 0.67$ for the proposed model); and third, the extracted value from the proposed model is a closer match to the intrinsic work of adhesion measured upon retraction.

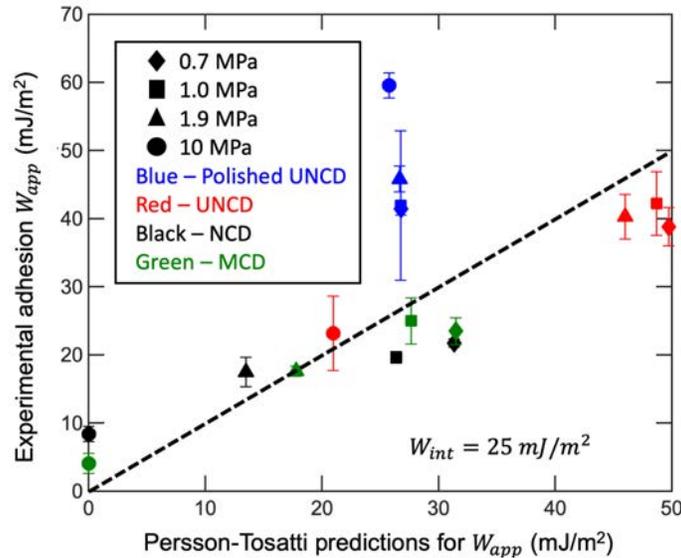

**Figure S8:** The experimental measurements of $W_{app}$ can be compared against the unmodified Persson-Tosatti model, which does not account for the change in area of the soft elastomer.

# References

1. Jacobs TDB, Junge T, Pastewka L (2017) Quantitative characterization of surface topography using spectral analysis. *Surf Topogr Metrol Prop* 5(1):013001.
2. Gujrati A, Khanal SR, Pastewka L, Jacobs TDB (2018) Combining TEM, AFM, and profilometry for quantitative topography characterization across all scales. *ACS Appl Mater Interfaces* 10(34):29169–29178.
3. Johnson KL, Kendall K, Roberts AD (1971) Surface energy and the contact of elastic solids. *Proc R Soc London Ser A-Mathematical Phys Eng Sci* 324(1558):301–313.
4. Krick BA, Vail JR, Persson BNJ, Sawyer WG (2012) Optical in situ micro tribometer for analysis of real contact area for contact mechanics, adhesion, and sliding experiments. *Tribol Lett* 45(1):185–194.
5. Persson BNJ, Tosatti E (2001) The effect of surface roughness on the adhesion of elastic solids. *J Chem Phys* 115(12):5597.
6. Ramisetti SB, et al. (2011) The autocorrelation function for island areas on self-affine surfaces. *J Phys Condens Matter* 23(21):215004.